\documentclass[pra,onecolumn,floatfix,a4paper,superscriptaddress]{revtex4}
\usepackage{bm,color,graphicx,amsmath,txfonts}

\usepackage[colorlinks, citecolor=blue,linkcolor=blue]{hyperref}

\usepackage{float}

\begin{document}

\title{Extracted work as measure of entanglement in optomechanics}

\author{Hamza Harraf}
\affiliation{LPHE-Modeling and Simulation, Faculty of Sciences, Mohammed V University in Rabat, Rabat, Morocco.}	
\author{M'bark Amghar} 
\affiliation{LPTHE-Department of Physics, Faculty of Sciences, Ibnou Zohr University, Agadir 80000, Morocco}
\author{Wiam Kaydi}
\affiliation{LPHE-Modeling and Simulation, Faculty of Sciences, Mohammed V University in Rabat, Rabat, Morocco.}
\author{Mohamed Amazioug} \thanks{amazioug@gmail.com}
\affiliation{LPTHE-Department of Physics, Faculty of Sciences, Ibnou Zohr University, Agadir 80000, Morocco}
\author{Rachid Ahl Laamara}
\affiliation{LPHE-Modeling and Simulation, Faculty of Sciences, Mohammed V University in Rabat, Rabat, Morocco.}
\affiliation{Centre of Physics and Mathematics, CPM, Faculty of Sciences, Mohammed V University in Rabat, Rabat, Morocco.}

\begin{abstract}
In this work, we investigate quantum entanglement and work extraction in two distinct optomechanical systems. The first system consists of two spatially separated Fabry-P\'erot cavities driven by squeezed light in the resolved-sideband regime, while the second system comprises a laser field incident on a vibrating mirror. We analyze the entanglement dynamics between optical and mechanical modes, as well as quantum correlations in a mixed optomechanical bipartite system (optic-optic mode) mediated by radiation pressure. Using logarithmic negativity as a measure, we quantify the entanglement evolution in both optic-optic and mirror-mirror bipartite subsystems. Furthermore, we show how three distinct types of extractable work vary with system parameters and examine their relationship with entanglement. Our results highlight the relationship between quantum correlations and extracted work in optomechanical system, offering insights for quantum information processing and energy transfer in hybrid systems.
\end{abstract}

\maketitle
\section{Introduction}

The study of the interaction of light with mechanical objects, including mirrors, through radiation pressure at extremely small sizes is known as cavity optomechanics \cite{intro_14_5, intro_15}. This field has enabled the development of innovative techniques for generating and manipulating squeezed states of light, which are a specific type of quantum state. Recently, cavity optomechanics has found significant applications in quantum information processing, including the creation of quantum entangled states \cite{Asadian16,Asjad16,Asjad18,Teklu18,amazioug20,amazioug2020,Abdi21,intro_16}, cooling mechanical modes to their quantum ground states \cite{intro_17}, photon blockade \cite{intro_18}, superconducting elements \cite{intro_17}, and interactions between massive mechanical oscillators \cite{intro_20}. Cavity optomechanics has emerged as an extension of the work in cavity quantum electrodynamics (CQED), which focuses on controlling the interaction between light (photons) and atoms at the quantum level. In CQED, the strong confinement of photons enables single quanta to exert a significant influence on atom-cavity dynamics in the strong coupling regime. Recent experimental advancements in achieving strong coupling have led to the observation of remarkable quantum phenomena, including quantum phase gates \cite{intro_21}, Fock state generation \cite{intro_22}, and quantum non demolition detection of single cavity photons \cite{intro_23}.

Entanglement a fundamental phenomenon in quantum information and communication is characterized by nonlocal quantum correlations between subsystems within an entangled state \cite{intro_1,28}. Recent years have witnessed growing interest in quantum state transfer within optomechanical systems \cite{intro_3}, with research focusing on entanglement generation between mechanical and optical modes \cite{Sete2014} as well as between reflected optical modes (denoted as optics 1 and optics 2) on movable mirrors \cite{intro_5}. Often termed quantum nonlocality \cite{intro_6}, entanglement lies at the heart of optomechanical systems \cite{intro_7}, enabling breakthroughs in precision measurements such as gravitational wave detection \cite{intro_8} and atomic force microscopy \cite{intro_9}. Radiation pressure, arising from an optical beam incident on a mirror, couples the mirror’s vibrational mode to the reflected optical sideband modes, facilitating the generation of continuous-variable (CV) entanglement. This approach offers experimental advantages over discrete-variable entanglement, as CV entanglement can be efficiently realized using linear optical elements \cite{intro_10}. Furthermore, entanglement serves as a critical benchmark for distinguishing quantum correlations from classical ones, as well as entangled states from separable ones \cite{intro_11}.

In this work, we investigate two distinct optomechanical systems to explore the interplay between quantum correlations and quantum thermodynamics \citep{33,Sete2014}. The first system comprises two identical Fabry-P\'erot cavities, each containing a fixed and a movable mirror. The second system features a vibrating mirror subjected to an intense quasi-monochromatic laser field. We employ logarithmic negativity to quantify entanglement and introduce an innovative approach: using extractable work as a diagnostic tool for detecting entanglement in optomechanical systems. Our results analyze how entanglement and extractable work vary with system parameters, as well as the influence of temperature on both measures.

This paper is organized as follows: In Section \ref{sec1}, we introduce systems 1 and 2, derive the quantum nonlinear Langevin equations (QLEs) from the Hamiltonian describing the optical and mechanical degrees of freedom, and compute the covariance matrices for the mirror-mirror and optic-optic subsystems. Section \ref{thermo} explores the connection between quantum entanglement and quantum thermodynamics in cavity optomechanical systems. In Section \ref{sys1}, we analyze the dependence of entanglement and extractable work on system parameters for both configurations. Finally, Section \ref{conc} presents our conclusions and outlook.

\section{Model and Hamiltonian} \label{sec1}
\subsection{First optomechanical system}

\begin{figure}[H]
\centering
\includegraphics[scale=0.6]{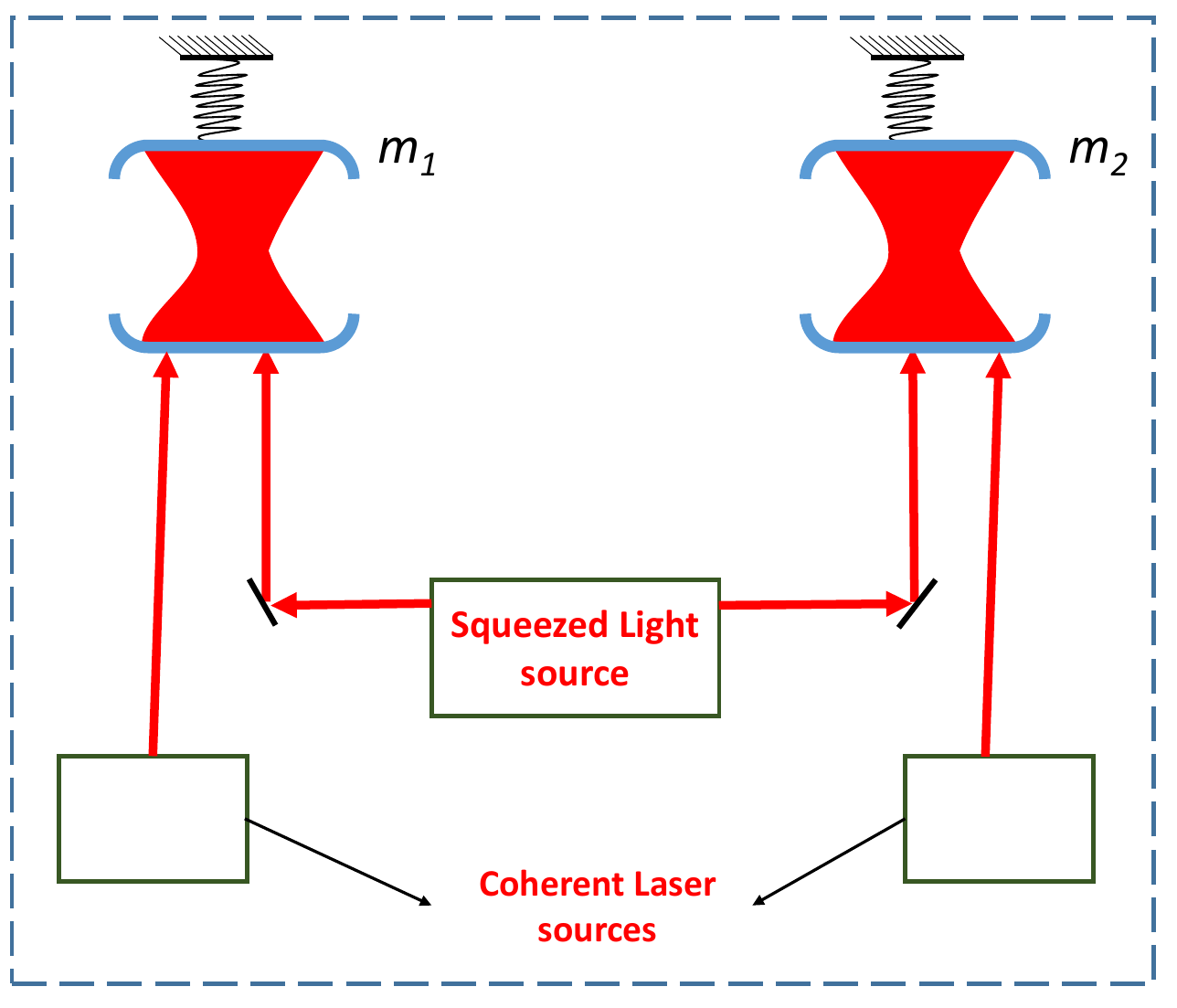} 
\caption{Schematic diagram of the coupled-cavity system. Two optomechanical cavities are driven by a common two-mode squeezed light source generated via spontaneous parametric down-conversion (SPDC), with each cavity additionally pumped by a coherent laser field of amplitude $\mathcal{O}$. Both cavities contain movable mirrors ($m_1$ and $m_2$) that couple to the intracavity field via radiation pressure.}
\label{fig_1}
\end{figure}

In this section (see Fig. \ref{fig_1}), our system consists of two Fabry-P\'erot cavities coupled to a common two-mode squeezed light source generated by parametric down-conversion. Each cavity contains one fixed mirror with finite transmission and one perfectly reflective movable mirror (denoted as quantum harmonic oscillators $m_1$ and $m_2$), with each cavity receiving a single mode of the squeezed vacuum field driven by a coherent laser pump. For simplicity, we hava the fixed mirrors as having finite transmissivity while assuming perfect reflectivity for the movable mirrors.\\
 The total Hamiltonian 
\begin{equation}\mathcal{H}_{T} = \mathcal{H}_1 + \mathcal{H}_2,
\end{equation}
describes our coupled optomechanical system, where for each cavity $i = 1,2$ in the rotating frame ($\hbar = 1$) \cite{28,Sete2014}, is given by:
\begin{align}\label{Ham}
\mathcal{H}_i &= \Omega_{m_i}d^{\dagger}_id_i + (\Omega_{c_i}-\Omega_{L_i})c^{\dagger}_ic_i \nonumber \\
&\quad + g_ic^{\dagger}_ic_i(d^{\dagger}_i+d_i) + (c^{\dagger}_i\mathcal{O}_ie^{i\phi_i} + c_i\mathcal{O}_ie^{-i\phi_i}).
\end{align}
Here, the first term represents the mechanical energy with mirror frequencies $\Omega_{m_i}$ and masses $m_i$, where the phonon operators $d_i,d^\dagger_i$ satisfy $[d_i,d^\dagger_i] = 1$. The second term gives the cavity field energy, with photon operators $c_i,c^\dagger_i$ ($[c_i,c^\dagger_i] = 1$) for cavities of frequencies $\Omega_{c_i}$, detuned from the laser frequencies $\Omega_{L_i}$. The third term describes radiation-pressure coupling with rates $g_i = \sqrt{\Omega^2_{c_i}L_i^{-2}\hbar/(m_i\Omega_{m_i})}$ for cavity lengths $L_i$ \cite{intro_15}. The last term represents laser driving with amplitudes $\mathcal{O}_i = \sqrt{2\kappa_iP_i/(\hbar\Omega_{L_i})}$, where $\kappa_i$ are cavity damping rates, $P_i$ the pump powers, and $\phi_i$ the input phases.
The dynamical evolution of the system is described by a set of nonlinear quantum Langevin equations \cite{mode1_c}
\begin{align}
\label{eq2}
\dot{d}_1=-\left(i\Omega_{m_1}+\frac{\Gamma_1}{2}\right)d_1-ig_1c_1^{\dagger}c_1+\sqrt{\Gamma_1}d^{in}_1,&\quad 
\dot{c}_1=-\left(\frac{\kappa_1}{2}-i\Delta_1\right)c_1-ig_1c_1(d^{\dagger}_1+d_1)-i\mathcal{O}_1e^{i\phi}+\sqrt{\kappa_1}c_1^{in}, \\ \nonumber
\dot{d}_2=-\left(i\Omega_{m_2}+\frac{\Gamma_2}{2}\right)d_2-ig_2c_2^{\dagger}c_2+\sqrt{\Gamma_2}d^{in}_2,&\quad
\dot{c}_2=-\left(\frac{\kappa_2}{2}-i\Delta_2\right)c_2-ig_2c_2(d^{\dagger}_2+d_2)-i\mathcal{O}_2e^{i\phi}+\sqrt{\kappa_2}c_2^{in}, 
\end{align}
with $\Delta_1 = \Omega_{L_1} - \Omega_{r_1}$($\Delta_2 = \Omega_{L_2} - \Omega_{r_2}$) being the laser detuning, {$\Gamma_1$($\Gamma_2$) the mechanical damping rate, $d^{in}_1$($d^{in}_2$) the noise operator characterizing the coupling between the mechanical mode and its environment, and $c^{in}_1$($c^{in}_2$) the squeezed vacuum operator.

For large values of the mechanical quality factor, we can assume the mechanical baths are Markovian. The non-zero correlation functions \cite{mode1_d} are given by
\begin{equation}
\langle d^{in}_1(t)d^{in\dagger}_1(t')\rangle=(n_1+1)\delta(t-t'), \quad \langle d^{in}_2(t)d^{in\dagger}_2(t')\rangle=(n_2+1)\delta(t-t'),
\end{equation}
\begin{equation}
\langle d^{in\dagger}_1(t)d^{in}_1(t')\rangle=n_1\delta(t-t'), \quad \langle d^{in\dagger}_2(t)d^{in}_2(t')\rangle=n_2\delta(t-t').
\end{equation}
here, $n_1 = \frac{1}{\exp\left(\frac{\hbar\Omega_{m_1}}{k_BT_1}\right)-1}$ and $n_2 = \frac{1}{\exp\left(\frac{\hbar\Omega_{m_2}}{k_BT_2}\right)-1}$ are the mean phonon numbers. The squeezed vacuum operators $c^{in}_1$ ($c^{in}_2$) and $c^{in\dagger}_1$ ($c^{in\dagger}_2$) exhibit non-zero correlation properties \cite{mode1_e}.
\begin{equation}
\langle c^{in}_1(t)c^{in\dagger}_1(t')\rangle=(N+1)\delta(t-t'),\quad \langle c^{in}_2(t)c^{in\dagger}_2(t')\rangle=(N+1)\delta(t-t').
\end{equation}
\begin{equation}
\langle c^{in\dagger}_1(t)c^{in}_1(t')\rangle=(N+1)\delta(t-t'),\quad \langle  c^{in\dagger}_2(t)c^{in}_2(t')\rangle=(N+1)\delta(t-t'),
\end{equation}
\begin{equation}
\langle c^{in}_1(t)c^{in}_2(t')\rangle=Me^{-\Omega_m(t+t')}\delta(t-t'),\quad \langle c^{in}_2(t)c^{in}_1(t')\rangle=Me^{-\Omega_m(t+t')}\delta(t-t'),
\end{equation}
\begin{equation}
\langle c^{in\dagger}_1(t)c^{in\dagger}_2(t')\rangle=Me^{\Omega_m(t+t')}\delta(t-t'),\quad \langle c^{in\dagger}_2(t)c^{in\dagger}_1(t')\rangle=Me^{\Omega_m(t+t')}\delta(t-t'),
\end{equation}
where, $N=\sinh2r$ and $M=\sinh r\cosh r$, where $r$ is the squeezing parameter that describes the squeezed light. For simplicity, we assume equal mechanical frequencies, i.e., $\Omega_{m_1}=\Omega_{m_2}=\Omega_m$.

\subsubsection{Linearization of quantum Langevin equations}

To linearize the quantum Langevin equations (LQLEs) (Eq. (\ref{eq2})), which are not generally solvable analytically, we employ the method described in reference \cite{mode1_f}
\begin{align}
\label{eq9}
\quad d_1=\langle d_1\rangle+\delta d_1,&\quad c_1=\langle c_1\rangle+\delta c_1,\\
 d_2=\langle d_2\rangle+\delta d_2,&\quad  c_2=\langle c_2\rangle+\delta c_2,\nonumber
\end{align}
with $\delta c_1$ ($\delta c_2$) and $\delta d_1$ ($\delta d_2$) are the fluctuation operators. The steady-state averages for operators $c_1$ ($c_2$) and $d_1$ ($d_2$) are $\langle c_1 \rangle$ ($\langle c_2 \rangle$) and $\langle d_1 \rangle$ ($\langle d_2 \rangle$), respectively.

Utilizing Eq. \eqref{eq2}, one can obtain the explicit expressions for $\langle c_1 \rangle$ ($\langle c_2 \rangle$) and $\langle d_1 \rangle$ ($\langle d_2 \rangle$)
\begin{align}
\label{eq10}
\langle d_1\rangle=\frac{-ig_1|\langle c_1\rangle|^2}{\frac{\Gamma_1}{2}-i\Omega_{m_1}},&\quad \langle c_1\rangle=\frac{-i\mathcal{O}_1e^{i\phi_1}}{\frac{k_1}{2}-i\Delta_1^{'}},\\
 \langle d_2\rangle=\frac{-ig_2|\langle c_2\rangle|^2}{\frac{\Gamma_2}{2}-i\Omega_{m_2}},&\quad \langle c_2\rangle=\frac{-i\mathcal{O}_2e^{i\phi_2}}{\frac{k_2}{2}-i\Delta_2^{'}},\nonumber
\end{align}
where, $\Delta_1'=\Delta_1-g_1\left(\langle d_1\rangle+\langle d_1\rangle^*\right)$ and $\Delta_2'=\Delta_2-g_2(\langle d_2\rangle+\langle d_2\rangle^*)$ represent the effective cavity detuning, accounting for the displacement of the mirrors caused by the radiation pressure force. By substituting Eq.~\eqref{eq9} into Eq.~\eqref{eq2}, one can derive the following
\begin{equation}
		\begin{aligned}
		\label{eq1_1}
		\delta\dot{d}_1=-(\frac{\Gamma_1}{2}+i\Omega_{m_1})\delta d_1+G_1(\delta c_1-\delta c^{\dagger}_1)+\sqrt{\Gamma_1}d^{in}_1,&\quad
		\delta\dot{c}_1=-(\frac{\kappa_1}{2}-i\Delta_{_1}^{'})\delta c_1+G_1(\delta d_1-\delta d^{\dagger}_1)+\sqrt{\kappa_1}c^{in}_1,\\ 
		\delta\dot{d}_2=-(\frac{\Gamma_2}{2}+i\Omega_{m_2})\delta d_2+G_2(\delta c_2-\delta c^{\dagger}_2)+\sqrt{\Gamma_2}d^{in}_2,&\quad
		\delta\dot{c}_2=-(\frac{\kappa_2}{2}-i\Delta_{_2}^{'})\delta c_2+G_2(\delta d_2-\delta d^{\dagger}_2)+\sqrt{\kappa_2}c^{in}_2,\\ 
	\end{aligned}
\end{equation}
where, $G_1=g_1|\langle c_1\rangle|$ and $G_2=g_2|\langle c_2\rangle|$ are the effective optomechanical coupling strengths. We have selected the input field phases such that $\phi_1=-\arctan\left(\frac{2\Delta_1'}{\kappa_1}\right)$ and $\phi_2=-\arctan\left(\frac{2\Delta_2'}{\kappa_2}\right)$. This choice implies that $\langle c_1\rangle=-i|\langle c_1\rangle|$ and $\langle c_2\rangle=-i|\langle c_2\rangle|$. Introducing notation for operators: 
\begin{align}
\label{eq12}
\delta \tilde{d}_1(t)=\delta d_1(t)e^{i\Omega_{m_1}t},&\quad   \delta \tilde{c}_1(t)=\delta c_1(t)e^{-i\Delta_1^{'}t},\\
 \delta \tilde{d}_2(t)=\delta d_2(t)e^{i\Omega_{m_2}t} ,&\quad \delta \tilde{c}_2(t)=\delta c_2(t)e^{-i\Delta_2^{'}t}.\nonumber
\end{align}
Eqs. \eqref{eq1_1} and \eqref{eq12} are equivalent to
\begin{equation}
	\begin{aligned}
		\label{eq13}
		\delta\dot{\tilde{d}}_1=-\frac{\Gamma_1}{2}\delta\tilde{d}_1+G_1(\delta\tilde{c}_1e^{i(\Delta_1^{'}+\Omega_{m_1})t}-\delta\tilde{c}_1^{\dagger}e^{-i(\Delta_1^{'}-\Omega_{m_1})t})+\sqrt{\Gamma_1}\tilde{d}^{in}_1,
		&\\
		\delta\dot{\tilde{c}}_1=-\frac{\kappa_1}{2}\delta\tilde{c}_1+G_1(\delta\tilde{d}^{\dagger}_1e^{-i(\Delta_1^{'}-\Omega_{m_1})t}+\delta\tilde{d}_1e^{-i(\Delta_1^{'}+\Omega_{m_1})t})+\sqrt{\kappa_1}\tilde{c}^{in}_1,\\ 
		\delta\dot{\tilde{d}}_2=-\frac{\Gamma_2}{2}\delta\tilde{d}_2+G_2(\delta\tilde{c}_2e^{i(\Delta_2^{'}+\Omega_{m_2})t}-\delta\tilde{c}_2^{\dagger}e^{-i(\Delta_2^{'}-\Omega_{m_2})t})+\sqrt{\Gamma_2}\tilde{d}^{in}_2,
		&\\
		\delta\dot{\tilde{c}}_2=-\frac{\kappa_2}{2}\delta\tilde{c}_2+G_2(\delta\tilde{d}^{\dagger}_2e^{-i(\Delta_2^{'}-\Omega_{m_2})t}+\delta\tilde{d}_2e^{-i(\Delta_2^{'}+\Omega_{m_2})t})+\sqrt{\kappa_2}\tilde{c}^{in}_2,\\ 
	\end{aligned}
\end{equation}
where
\begin{equation}
	\begin{aligned}
		\tilde{d}_1^{in}=e^{i\Omega_{m_1}t}d^{in}_1,&\quad \tilde{c}_1^{in}=e^{-i\Delta_{1}^{'}t}c^{in}_1,\\
		\tilde{d}_2^{in}=e^{i\Omega_{m_2}t}d^{in}_2,&\quad \tilde{c}_2^{in}=e^{-i\Delta_{2}^{'}t}c^{in}_2.
	\end{aligned}
\end{equation}
In the rotating wave approximation (RWA) \cite{intro_15}, where $\Omega_{m_1} \gg \kappa_1$ and $\Omega_{m_2} \gg \kappa_2$, terms oscillating at $\pm 2\Omega_m$ can be disregarded. This allows us to approximate $\Delta_1' \approx \Delta_1$ and $\Delta_2' \approx \Delta_2$. When the cavity is driven at the red sideband, with $\Delta_1' = -\Omega_{m_1}$ and $\Delta_2' = -\Omega_{m_2}$, Eq.~\eqref{eq13} simplifies to
\begin{equation}
	\begin{aligned}
		\label{eq_15}
		\delta\dot{\tilde{d}}_1=-\frac{\Gamma_1}{2}\delta\tilde{d}_1+G_1\delta\tilde{c}_1+\sqrt{\Gamma_1}\tilde{d}^{in}_1,&\quad
		\delta\dot{\tilde{c}}_1=-\frac{\kappa_2}{2}\delta\tilde{c}_1+G_1\delta\tilde{d}_1+\sqrt{\kappa_1}\tilde{c}^{in}_1,\\ 
		\delta\dot{\tilde{d}}_2=-\frac{\Gamma_2}{2}\delta\tilde{d}_2+G_2\delta\tilde{c}_2+\sqrt{\Gamma_2}\tilde{d}^{in}_2,&\quad
		\delta\dot{\tilde{c}}_2=-\frac{\kappa_1}{2}\delta\tilde{c}_2+G_2\delta\tilde{d}_2+\sqrt{\kappa_2}\tilde{c}^{in}_2.
	\end{aligned}
\end{equation}

\subsubsection{Covariance matrix}

For optimal quantum-state transfer from the two modes of squeezed vacuum to the optical modes, the covariance matrix (CM) formalism is essential for quantifying correlations between the mechanical and optical modes, as it characterizes the evolution of the system's steady state.
We assume that the external laser drives have the same strength and that the temperatures of the thermal baths of the two mechanical mirrors are equal, i.e., $n_1=n_2=n$ and $T_1=T_2=T$. Additionally, we set the following parameters to be equal for both cavities: $m_1=m_2=m$, $\Omega_{r_1}=\Omega_{r_2}=\Omega_r$, $\Omega_{m_1}=\Omega_{m_2}=\Omega_m$, $\kappa_1=\kappa_2=\kappa$, and $\Gamma_1=\Gamma_2=\Gamma$.
We present the EPR-type quadrature operators for the two subsystems to derive the explicit formula for the CM
\begin{equation}
	\begin{aligned}
		\delta\tilde{X}_{d_1}=\frac{\delta\tilde{d}^{\dagger}_1+\delta\tilde{d}_1}{\sqrt{2}},&\quad \quad \delta\tilde{Y}_{d_1}=\frac{\delta\tilde{d}_1-\delta\tilde{d}^{\dagger}_1}{\sqrt{2}},\\ 
		\delta\tilde{X}_{d_2}=\frac{\delta\tilde{d}^{\dagger}_2+\delta\tilde{d}_2}{\sqrt{2}},&\quad \quad \delta\tilde{Y}_{d_2}=\frac{\delta\tilde{d}_2-\delta\tilde{d}^{\dagger}_2}{\sqrt{2}},\\ 
		\delta\tilde{X}_{c_1}=\frac{\delta\tilde{c}^{\dagger}_1+\delta\tilde{c}_1}{\sqrt{2}},&\quad  \quad \delta\tilde{Y}_{c_1}=\frac{\delta\tilde{c}_1-\delta\tilde{c}^{\dagger}_1}{\sqrt{2}},\\ 
		\delta\tilde{X}_{c_2}=\frac{\delta\tilde{c}^{\dagger}_2+\delta\tilde{c}_2}{\sqrt{2}},&\quad  \quad \delta\tilde{Y}_{c_2}=\frac{\delta\tilde{c}_2-\delta\tilde{c}^{\dagger}_2}{\sqrt{2}}.
	\end{aligned}
\end{equation}
Consequently, it is convenient to rewrite Eq. \eqref{eq_15} in terms of the quadrature operators
\begin{align}
\label{eq_18}
\delta\dot{\tilde{X}}_{d_1}=-\frac{\Gamma}{2}\delta\tilde{X}_{d_1}+G\delta\tilde{X}_{c_1}+\sqrt{\Gamma}\tilde{X}^{in}_{d_1},&\quad \quad 
\delta\dot{\tilde{X}}_{d_2}=-\frac{\Gamma}{2}\delta\tilde{X}_{d_2}+G\delta\tilde{X}_{c_2}+\sqrt{\Gamma}\tilde{X}^{in}_{d_2},\\ 
\delta\dot{\tilde{Y}}_{d_1}=-\frac{\Gamma}{2}\delta\tilde{Y}_{d_1}+G\delta\tilde{Y}_{c_1}+\sqrt{\Gamma}\tilde{Y}^{in}_{d_1},& \quad \quad 
\delta\dot{\tilde{Y}}_{d_2}=-\frac{\Gamma}{2}\delta\tilde{Y}_{d_2}+G\delta\tilde{Y}_{c_2}+\sqrt{\Gamma}\tilde{Y}^{in}_{d_2},\\ 
\delta\dot{\tilde{X}}_{c_1}=-\frac{\kappa}{2}\delta\tilde{X}_{c_1}+G\delta\tilde{X}_{d_1}+\sqrt{\kappa}\tilde{X}^{in}_{c_1},& \quad \quad
\delta\dot{\tilde{X}}_{c_2}=-\frac{\kappa}{2}\delta\tilde{X}_{c_2}+G\delta\tilde{X}_{d_2}+\sqrt{\kappa}\tilde{X}^{in}_{c_2},\\
\label{eq_23} 
\delta\dot{\tilde{Y}}_{c_1}=-\frac{\kappa}{2}\delta\tilde{Y}_{c_1}+G\delta\tilde{Y}_{d_1}+\sqrt{\kappa}\tilde{Y}^{in}_{c_1},& \quad \quad
\delta\dot{\tilde{Y}}_{c_2}=-\frac{\kappa}{2}\delta\tilde{Y}_{c_2}+G\delta\tilde{Y}_{d_2}+\sqrt{\kappa}\tilde{Y}^{in}_{c_2}. 
\end{align}
With
\begin{equation}
	\begin{aligned}
		\tilde{X}_{d_1}^{in}=\frac{\tilde{d}^{in\dagger}_1+\tilde{d}^{in}_1}{\sqrt{2}},&\quad \quad \quad \tilde{Y}_{d_1}^{in}=\frac{\tilde{d}_1^{in}-\tilde{d}^{in\dagger}_1}{\sqrt{2}},\\ 
		\tilde{X}_{d_2}^{in}=\frac{\tilde{d}^{in\dagger}_2+\tilde{d}^{in}_2}{\sqrt{2}},&\quad \quad \quad \tilde{Y}_{d_2}^{in}=\frac{\tilde{d}_2^{in}-\tilde{d}^{in\dagger}_2}{\sqrt{2}},\\ 
		\tilde{X}_{c_1}^{in}=\frac{\tilde{c}^{in\dagger}_1+\tilde{c}^{in}_1}{\sqrt{2}},&\quad \quad \quad \tilde{Y}_{c_1}^{in}=\frac{\tilde{c}_1^{in}-\tilde{c}^{in\dagger}_1}{\sqrt{2}}\\ 
		\tilde{X}_{c_2}^{in}=\frac{\tilde{c}^{in\dagger}_2+\tilde{c}^{in}_2}{\sqrt{2}},&\quad \quad \quad \tilde{Y}_{c_2}^{in}=\frac{\tilde{c}_2^{in}-\tilde{\mathsf{c}}^{in\dagger}_2}{\sqrt{2}}\\ 
	\end{aligned}
\end{equation}
Eqs.\eqref{eq_18}–\eqref{eq_23} can be written in a compact matrix form \cite{mode1_h}:
\begin{equation}
\dot{\lambda}(t)=\mathcal{A}\lambda(t)+\mathsf{f}(t)
\end{equation}
with \begin{align}
\lambda(t)=&[\delta \tilde{X}_{d_1}(t),\delta \tilde{Y}_{d_1}(t),\delta \tilde{X}_{d_2}(t),\delta \tilde{Y}_{d_2}(t),\delta \tilde{X}_{c_1}(t),\delta \tilde{Y}_{c_1}(t),\delta \tilde{X}_{c_2}(t),\delta \tilde{Y}_{c_2}(t)]^T,\\ 
\mathsf{f}(t)=&[\delta \tilde{X}^{in}_{d_1}(t),\delta \tilde{Y}^{in}_{d_1}(t),\delta \tilde{X}^{in}_{d_2}(t),\delta \tilde{Y}^{in}_{d_2}(t),\delta \tilde{X}^{in}_{c_1}(t),\delta \tilde{Y}^{in}_{c_1}(t),\delta \tilde{X}^{in}_{c_2}(t),\delta \tilde{Y}^{in}_{c_2}(t)]^T,\nonumber
\end{align}
and the drift matrix $\mathcal{A}$ is given by
\begin{equation}
\mathcal{A}=
\begin{pmatrix}
-\frac{\Gamma}{2}&0&0&0&G&0&0&0\\
0&-\frac{\Gamma}{2}&0&0&0&G&0&0\\
0&0&-\frac{\Gamma}{2}&0&0&0&G&0\\
0&0&0&-\frac{\Gamma}{2}&0&0&0&G\\
-G&0&0&0&-\frac{\kappa}{2}&0&0&0\\
0&-G&0&0&0&-\frac{\kappa}{2}&0&0\\
0&0&-G&0&0&0&-\frac{\kappa}{2}&0\\
0&0&0&-G&0&0&0&-\frac{\kappa}{2}
\end{pmatrix}
\end{equation}
The system is stable; all eigenvalues of the drift matrix $\mathcal{A}$ are negative, given the condition $G > \kappa, \Gamma$ (i.e., $2G > \kappa + \Gamma$) \cite{intro_15}. The correlation matrix, $\mathcal{V}_{\alpha\beta}=\frac{\langle u_{\alpha}(t)u_{\beta}(t')+u_{\alpha}(t')u_{\beta}(t)\rangle}{2}$, can be used to characterize the steady state of quantum fluctuations in a bipartite Gaussian system. The covariance matrix can then be derived using the following Lyapunov equation \cite{mode1_P}
\begin{equation}
\mathcal{A}\mathcal{V}+\mathcal{V} \mathcal{A}^T=-\mathcal{D},
\end{equation}
where $\mathcal{D}$ is the matrix of stationary noise correlation functions, given by $\mathcal{D}_{\alpha\beta}=\frac{\langle n_{\alpha}(t)n_{\beta}(t')+n_{\alpha}(t')n_{\beta}(t)\rangle}{2\delta(t-t')}$,
\begin{equation}
\mathcal{D}=
\begin{pmatrix}
\Gamma(n+\frac{1}{2})&0&0&0&0&0&0&0\\
0&\Gamma(n+\frac{1}{2})&0&0&0&0&0&0\\
0&0&\Gamma(n+\frac{1}{2})&0&0&0&0&0\\
0&0&0&\Gamma(n+\frac{1}{2})&0&0&0&0\\
0&0&0&0&\kappa(N+\frac{1}{2})&0&M\kappa&0\\
0&0&0&0&0&\kappa(N+\frac{1}{2})&0&M\kappa\\
0&0&0&0&-M\kappa&0&\kappa(N+\frac{1}{2})&0\\
0&0&0&0&0&-M\kappa&0&\kappa(N+\frac{1}{2})
\end{pmatrix}.
\end{equation}
The steady-state covariance matrix of our system is explicitly given by
\begin{equation}
\mathcal{V}=
\begin{pmatrix}
\mathcal{V}_{m_1m_2}&\mathcal{V}_{mo}\\
\mathcal{V}_{mo}^T&\mathcal{V}_{o_1o_2}
\end{pmatrix}.
\end{equation}
The covariance matrices for each subsystem, namely $m_1m_2$ (mirror 1 and mirror 2 modes) and $o_1o_2$ (optical 1 and optical 2 modes), are given by
\begin{align}
\mathcal{V}_{m_1m_2}=
\begin{pmatrix}
v_{11}&0&v_{13}&0\\
0&v_{11}&0&-v_{13}\\
v_{13}&0&v_{11}&0\\
0&-v_{13}&0&v_{11}\\
\end{pmatrix},\quad \quad &
\mathcal{V}_{o_1o_2}=
\begin{pmatrix}
v_{22}&0&v_{57}&0\\
0&v_{22}&0&-v_{57}\\
v_{57}&0&v_{22}&0\\
0&-v_{57}&0&v_{22}\\
\end{pmatrix}
,
\end{align}

\begin{align}
v_{11}=\frac{\kappa C\cosh(2r)+(1+2n)(\kappa+\Gamma(1+C))}{2(\kappa+\Gamma)(1+C)},\quad \quad &
v_{13}=\frac{\kappa C\sinh(2r)}{2(\kappa+\Gamma)(1+C)},\\ \nonumber
v_{22}=\frac{(\Gamma+\kappa(1+ C))\cosh(2r)+(1+2n)\Gamma C}{2(\kappa+\Gamma)(1+C)},\quad \quad & v_{57}=\frac{(\Gamma+\kappa(1+ C))\sinh(2r)}{2(\kappa+\Gamma)(1+C)}.\nonumber
\end{align}
These elements are precisely consistent with the findings in references \cite{28,Sete2014}. Additionally, the optomechanical cooperativity is defined as in \cite{intro_15}
\begin{equation}
C=\frac{4G^{2}}{\Gamma \kappa}=\frac{8\Omega_c^2P}{m\Gamma\Omega_m\Omega_LL^2[(\frac{\kappa}{2})^2+\Omega_m^2]}.
\end{equation}

\subsection{Second optomechanical system} \label{sec2}

The annihilation operator $d$ describes a movable mirror (mechanical mode), while $c_1$ and $c_2$ describe two weak inelastic sideband modes (Stokes and anti-Stokes annihilation operators, respectively). These operators satisfy the usual commutation relations $[c_1,c_1^{\dagger}]=1$ and $[c_2,c_2^{\dagger}]=1$. Hamiltonian of this system is given by 
\begin{equation}
\label{eq11}
	H_{\mathrm{eff}}=-ig_1(c_1d-c_1^{\dagger}d^{\dagger})-ig_2(c_2d^{\dagger}-c_2^{\dagger}d),
\end{equation} 
where, $g_1$ and $g_2$ are coupling constants proportional to $\sqrt{P}$, where $P$ represents the incident laser power. Equation \eqref{eq11} consists of two interaction terms: the first, between modes $c_1$ and $d$, is a parametric-type interaction that can lead to phase-space squeezing \cite{mode1_z}. The second interaction, between modes $c_2$ and $d$, is a beam-splitter interaction \cite{mode1_z}. Our goal is to determine the evolution of the system's covariance matrix.\\
\begin{figure}[!h]
\label{fig_2}
\centering
\includegraphics[scale=0.3]{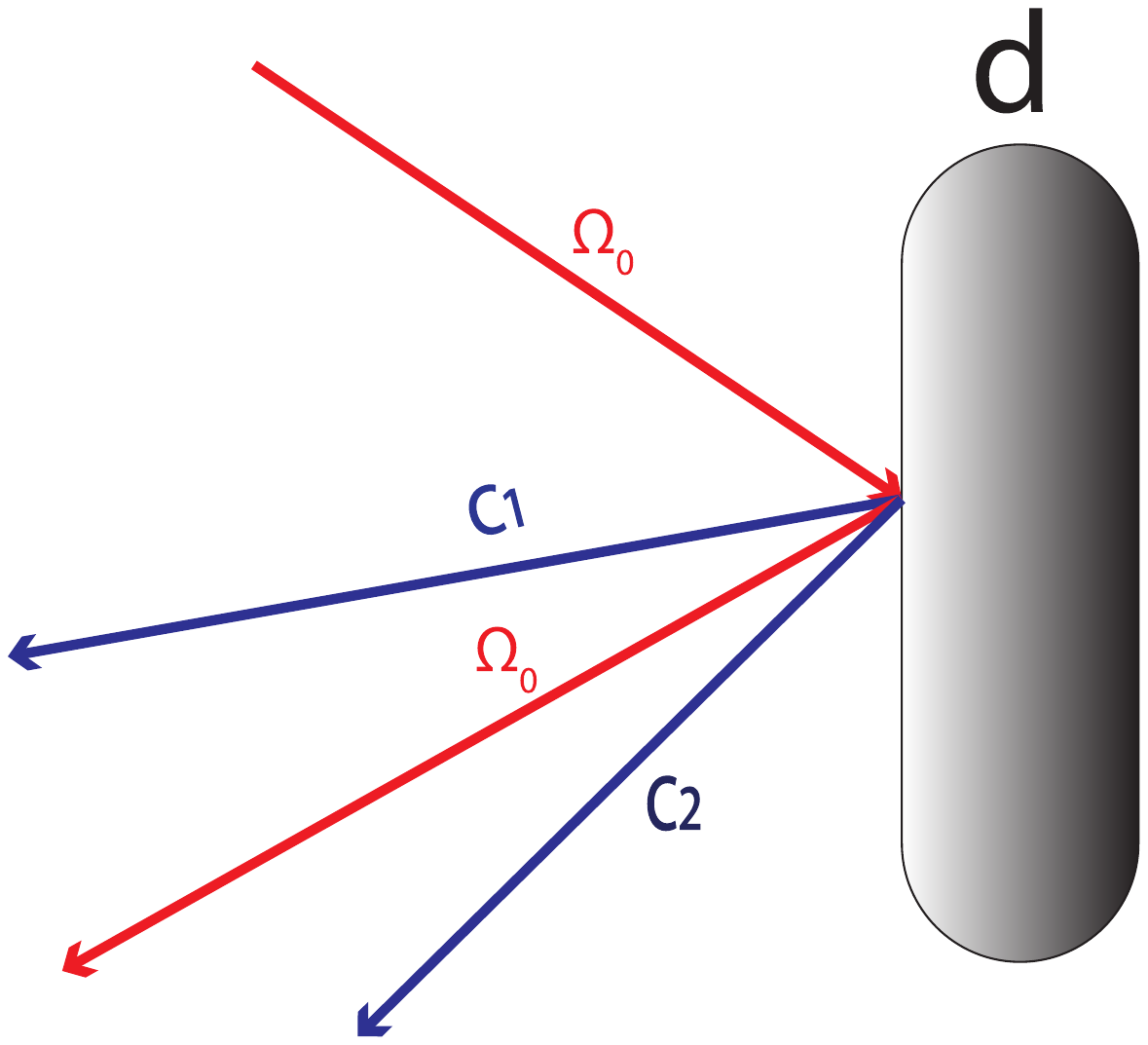} 
\caption{This schematic diagram illustrates an optomechanical system. A laser field, incident at frequency $\Omega_0$, interacts with the mechanical mode (mirror) vibrating at frequency $\Omega$. This interaction excites two sideband modes in the reflected field, occurring at frequencies $\Omega_1=\Omega_0-\Omega$ and $\Omega_2=\Omega_0+\Omega$.}
\end{figure}
The Hamiltonian presented in Eq.~\eqref{eq11} yields a set of linear Heisenberg
\begin{center}
 \begin{align}
 \frac{dc_1}{dt}=g_1d^{\dagger},& \quad\quad\quad\quad\quad\quad\quad\quad \frac{dc_2}{dt}=g_2d,& \quad\frac{dd}{dt}=g_1 c^{\dagger}_1-g_2 c_2.
 \end{align}
\end{center}
The solution to these equations is presented below
	\begin{align}
	c_1(t)=k_1c_1(0)+l_1c^{\dagger}_2(0)+l_2d^{\dagger}(0),&\quad c_2(t)=-l_1c_1^{\dagger}(0)+k_2c_2(0)+l_3d(0),&\quad d(t)=l_2c_1^{\dagger}(0)+l_3c_2(0)+k_3d(0),
	\end{align}
\begin{equation}
		\begin{aligned}
		k_1=\frac{[g_2^2-g_1^2\cos(\Omega t)]}{\Omega^2},&\quad l_1=\frac{g_1g_2[\cos(\Omega t)-1]}{\Omega^2},\\ 
		k_2=\frac{[g_2^2\cos(\Omega t)-g_1^2]}{\Omega^2},&\quad l_2=\frac{g_1\sin(\Omega t)}{\Omega},\\ 
		k_3=\cos(\Omega t),&\quad l_3=\frac{g_2\sin(\Omega t)}{\Omega}.	
	\end{aligned}
\end{equation}
Let $\Omega^2=g_2^2-g_1^2$ and $x=\frac{g_2}{g_1}$. The quadrature phase operators are expressed as $\chi=(P_1,X_1,P_2,X_2,P_d,X_d)$, where
\begin{equation}
		\begin{aligned}
		X_1=\frac{c_1+c_1^{\dagger}}{\sqrt{2}},&\quad\quad P_1=\frac{c_1-c_1^{\dagger}}{i\sqrt{2}},\\
		X_2=\frac{c_2+c_2^{\dagger}}{\sqrt{2}},&\quad\quad P_2=\frac{c_2-c_2^{\dagger}}{i\sqrt{2}},\\ 
		X_{d}=\frac{d+d^{\dagger}}{\sqrt{2}},&\quad\quad P_{d}=\frac{d-d^{\dagger}}{i\sqrt{2}}.
	\end{aligned}
\end{equation}
The explicit expressions for the position and momentum operators are as follows
\begin{equation}
		\begin{aligned}
		X_1(t)=k_1X_1(0)+l_1X_2(0)+l_2X_{d}(0),&\quad X_2(t)=-l_1X_1(0)+k_2X_2(0)+l_3X_{d}(0),& \quad X_{d}(t)=l_2X_1(0)+l_3X_2(0)+k_3X_{d}(0),\\
		P_1(t)=k_1P_1(0)-l_1P_2(0)-l_2P_{d}(0),&\quad P_2(t)=l_1P_1(0)+k_2P_2(0)+l_3P_{d}(0), & \quad  P_{d}(t)=-l_2P_1(0)-l_3P_2(0)+k_3P_{d}(0). 
	\end{aligned}
\end{equation}
Thus, the system's covariance matrix (CM) \cite{intro_15} is 
\begin{equation}
\mathcal{V}_{\alpha\beta}=\frac{\langle \chi_{\alpha}\chi_{\beta}+\chi_{\beta}\chi_{\alpha}\rangle}{2}-\langle \chi_{\alpha}\rangle\langle\chi_{\beta}\rangle,
\end{equation}
where, $\chi_{\alpha}$ is the six-dimensional vector of phase-space coordinates $(P_1,X_1,P_2,X_2,P_d,X_d)$. We consider a pure tripartite system initially in the vacuum state \cite{27}, where the variance is $V(\chi_{\alpha},0)=\langle\chi_{\alpha}(0)^2\rangle=1$, the mean is $\langle\chi_{\alpha}(0)\rangle=0$, and the cross-correlation is $\langle X_{\alpha}(0)P_{\beta}(0)\rangle=0$ (for $\alpha,\beta\in\{1,2,d\}$). We will use the following equations to examine the components of the CM
\begin{align}
\nonumber
	v_{11}=\frac{\langle (X_1(t))^2\rangle}{2}=k_1^2+l_1^2+l_2^2,&\quad v_{32}=\frac{\langle X_{d}X_2+X_2X_{d}\rangle}{2}=-l_1l_2-k_2l_3+l_3k_3, \\ \nonumber
	v_{33}=\frac{\langle (X_2(t))^2\rangle}{2}=l_1^2+k_1^2+l_3^2,&\quad v_{21}=\frac{\langle X_1X_2+X_2X_1\rangle}{2}=-k_1l_1+l_1k_2+k_2k_3, \\ 
	v_{22}=\frac{\langle (P_1(t))^2\rangle}{2}=k_1^2+l_1^2+l_2^2,& \quad v_{31}=\frac{\langle X_1X_{d}+X_{d}X_1\rangle}{2}=k_1l_2-l_1l_3+l_2k_3,
\end{align}
with $X_1X_2=X_2X_1$, $X_2X_d=X_dX_2$, and $X_1X_d=X_dX_1$.
\begin{align}
	v_{11}=&\frac{1}{(x^2-1)^2}\left[x^4+(1+\cos^2(\Omega t)-4\cos(\Omega t))\mathsf{x}^2+\cos^2(\Omega t)+(x^2-1)\sin^2(\Omega t)\right],\\
	v_{22}=&\frac{1}{(x^2-1)^2}\left[x^4\cos^2(\Omega t)+(1+\cos^2(\Omega t)-4\cos(\Omega t))x^2+1+\mathsf{x}^2(x^2-1)\sin^2(\Omega t)\right],
\end{align}
\begin{align}
	 v_{33}=\frac{(1+x^2)\sin^2(\Omega t)}{x^2-1}+\cos^2(\Omega t),&\quad v_{21}=\frac{x(1+x^2)[1-\cos(\Omega t)]^2}{(x^2-1)^2}+\frac{x\sin^2(\Omega t)}{x^2-1}.
\end{align}
\begin{align}
	 v_{31}=&\frac{1}{\sqrt{x^2-1}}\left(\frac{2x^2\sin(\Omega t)-(1+x^2)\sin(\Omega t)\cos(\Omega t)}{x^2-1}+\sin(\Omega t)\cos(\Omega t)\right),\\
	 v_{32}=&\frac{1}{\sqrt{x^2-1}}\left(\frac{2x\sin(\Omega t)-(1+x^2)\mathsf{x}\sin(\Omega t)\cos(\Omega t)}{x^2-1}+x\sin(\Omega t)\cos(\Omega t)\right),
	 \end{align}

	\begin{align}
	v_{12}=\frac{\langle P_1P_2+P_2P_1\rangle}{2}=-v_{21},&\quad v_{13}=\frac{\langle P_1P_{d}+P_{d}P_1\rangle}{2}=-v_{31},& v_{23}=\frac{\langle P_{d}P_2+P_2P_{d}\rangle}{2}=v_{32}. 
	\end{align}
We can thus write the pure tripartite system's covariance matrix as
\begin{equation}
	\mathcal{V}=
	\begin{pmatrix}
	v_{11}&0&v_{12}&0&v_{13}&0\\
	0&v_{11}&0&-v_{12}&0&-v_{13}\\
	v_{12}&0&v_{22}&0&v_{23}&0\\
	0&-v_{12}&0&v_{22}&0&v_{23}\\
	v_{13}&0&v_{23}&0&v_{33}&0\\
	0&-v_{13}&0&v_{23}&0&v_{33}\\
	\end{pmatrix}.
\end{equation}
Then, the covariance matrix of the optical modes 1 and 2 (subsystem $o_1o_2$) is written as
\begin{equation}
	\mathcal{V}_{12}(t)=
	\begin{pmatrix}
	v_{11}&0&v_{12}&0&\\
	0&v_{11}&0&-v_{12}&\\
	v_{12}&0&v_{22}&0\\
	0&-v_{12}&0&v_{22}&\\
	\end{pmatrix}.
\end{equation}
	
\section{Quantum entanglement and quantum thermodynamics}
	
This section focuses on two Gaussian demons, Alice and Bob, sharing a bipartite Gaussian state. Alice controls  mode $a$, and Bob controls mechanical mode $b$.

\subsection{Logarithmic negativity}
	
The covariance matrix of the two optomechanical systems simplifies to 
\begin{equation}
\mathcal{V}_{AB}=
\begin{pmatrix}
\mathcal{X}_{a}&\mathcal{Z}_{ab}\\
\mathcal{Z}_{ab}^{T}&\mathcal{Y}_{b}
\end{pmatrix}
\equiv
\begin{pmatrix}
	x&0&z&0&\\
	0&x&0&-z&\\
	\mathsf{z}&0&y&0\\
	0&-z&0&y&\\
	\end{pmatrix}.
\end{equation}
Covariance matrix describes the correlation between two modes of each subsystem. For first optomechanical system, we define $x=v_{11}$ (or $v_{22}$), and $z=v_{13}$ (or $v_{57}$). For second optomechanical system, we define $x=v_{11}$, $y=v_{22}$, and $z=v_{12}$. In the continuous-variable (CV) case, the logarithmic negativity $L_N$ is given by \cite{29,Adesso04,Plenio05}
\begin{equation}
\label{log}
L_{N}=\text{max}[0,-\ln(2\vartheta^{-})],
\end{equation}
where $\vartheta^{-}=\frac{1}{\sqrt{2}}\sqrt{\Lambda-\sqrt{\Lambda^{2}-4\det \mathcal{V}_{AB}}}$, is the minimum simplyctic eigenvalues of covariance matrix $\mathcal{V}_{AB}$ and the symbol $\Lambda$  is written as
\begin{equation}
\Lambda=\det \mathcal{X}_{a}+\det \mathcal{Y}_{b}-2\det \mathcal{Z}_{ab}.
\end{equation}
the entanglement between two modes exist only when $\vartheta^{-}<\frac{1}{2}$ \cite{30}. Next, we will use the logarithmic negativity to examine the steady and dynamical entanglement that exists between the two modes.

\subsection{Quantum thermodynamics} \label{thermo}
\subsubsection{One mode Perform Gaussian measurement}

We consider two systems Alice ($A$) is affected when Bob ($B$) performs a Gaussian measurement on the portion of the system assigned to him. The description of this measurement is provided in \citep{Brunelli17,33}
\begin{equation}
\mathcal{N}_{b}(\varepsilon)=\frac{1}{\pi}\Delta_{b}(\varepsilon)\rho^{N_{b}}\Delta_{b}^{\dagger}(\varepsilon),
\end{equation}
where $\Delta_b(\varepsilon)=\exp(\varepsilon\delta b^{\dagger}-\varepsilon^{\star}\delta b)$ is the displacement operator. $\rho^{N_b}$ is a pure Gaussian state with a vanishing first moment, and its covariance matrix is given by
\begin{equation}
\mathcal{C}^{N_{b}}=\dfrac{1}{2}\mathcal{R}(\mathcal{\varsigma})\mathrm{diag}(\upsilon,\upsilon^{-1})\mathcal{R}(\mathcal{\varsigma})^T,
\end{equation}
where, $\upsilon$ is a positive real number, and $\mathcal{R}(\varsigma)=\begin{pmatrix} \cos\varsigma &-\sin\varsigma\\ \sin\varsigma &\cos\varsigma \end{pmatrix}$ is a rotation matrix. Homodyne detection is represented by $\upsilon=0$, while heterodyne detection is represented by $\upsilon=1$. The outcome $\varepsilon$ Bob obtains from his measurement does not affect the state of Alice mode $\delta a$; thus, $\mathcal{X}^{N_b}_{a|\varepsilon}=\mathcal{X}_a^{N_b}$. The covariance matrix of Alice's constrained mode can be explicitly expressed as
\begin{equation}
\label{eq36}
\mathcal{X}_{a}^{N_{b}}=\mathcal{X}_{a}-\mathcal{Z}_{ab}(\mathcal{X}_{a}+\mathcal{C}^{N_{b}})^{-1}\mathcal{Z}_{ab}^{T}. 
\end{equation}
Bob's measurement pushes the state of mode $A$ out of equilibrium. However, through interaction with a heat bath over a long period, mode $A$ eventually returns to an equilibrium state $\mathcal{X}_a^{eq}$. Its average entropy is $\int \mathcal{S}(\mathcal{X}_{a|\varepsilon}^{N_b}) d\varepsilon p_{\varepsilon}=\mathcal{S}(\mathcal{X}_a^{N_b})$, because her state is not impacted by the measurement result. Alice is able to remove work from a nearby heat bath
\begin{equation}
\mathcal{W}=k_BT\left[\mathcal{S}(\mathcal{X}_{a}^{eq})-\mathcal{S}(\mathcal{X}_{a}^{N_{b}})\right].
\end{equation}
We use the example of the covariance matrix for a squeezed thermal state, where $\mathcal{X}_a^{eq}=\mathcal{X}_a$. The entropy of the covariance matrix is quantified using the second-order R\'enyi entropy $\mathcal{S}(y)=-\ln(Try^2)$. In the case of two modes Gaussian states, writes as 
\begin{equation}
\mathcal{S}_2(\mathcal{V}_{AB})=\frac{1}{2}\ln(\det \mathcal{V}_{AB}),
\end{equation}
the extracted work, is now given by \cite{Brunelli17}
\begin{equation}
\mathcal{W}^{(\upsilon)}=\dfrac{k_BT}{2}\ln\left(\frac{\det \mathcal{X}_{a}}{\det \mathcal{X}_{a}^{N_{b}}}\right)~, \quad \upsilon=0~,~1.
\end{equation}
The extractable work for homodyne ($\upsilon$ = 0) and heterodyne ($\upsilon=1$) writes as
\begin{equation}
\mathcal{W}^{(0)}=\frac{k_BT}{2}\ln\left(\frac{xy}{xy-z^2}\right),\quad
\mathcal{W}^{(0)}_{\mathrm{Sep}}=\frac{k_BT}{2}\ln\left(\frac{4xy}{2x+2y-1}\right),
\quad \mathcal{W}^{(0)}_{\mathrm{Max}}=\frac{k_BT}{2}\ln\left(\frac{4xy}{1-2|x-y|}\right),
\end{equation}
\begin{equation}
\quad \mathcal{W}^{(1)}=k_BT \ln\left(\frac{2xy+x}{2xy+\mathsf{x}-2z^2}\right),\quad
\mathcal{W}^{(1)}_{\mathrm{Sep}}=k_BT \ln\left(\frac{2x(2y+1)}{4x+2y-1}\right),\quad
\mathcal{W}^{(1)}_{\mathrm{Max}}=
\begin{cases}
    \frac{k_BT}{2}\ln 2x \quad \quad \text{if}\quad x \leq y\\
    k_BT \ln\left(\frac{2x(2y+1)}{4x-2y+1}\right) \quad \text{otherwise}
\end{cases}.
\end{equation}

\subsubsection{Two modes Perform Gaussian measurement}

When Bob and Alice both perform Gaussian measurements on the system's reduced state, the situation can be explained by
\begin{equation}
\mathcal{N}_{a}(\varepsilon)=\frac{1}{\pi}\Delta_{a}(\eta)\rho^{N_{a}}\Delta_{a}^{\dagger}(\eta),
\end{equation}
where the displacement operator is $\Delta_a(\eta)=\exp(\eta\delta a^{\dagger}-\eta^{\star}\delta a)$. $\rho^{N_a}$ is a pure Gaussian state with a vanishing first moment, and its covariance matrix is provided by
\begin{equation}
\mathcal{C}^{N_{b}}=\dfrac{\mathcal{R}(\mathcal{\varsigma})\mathrm{diag}(\iota,\iota^{-1})\mathcal{R}(\mathcal{\varsigma})^T}{2},
\end{equation}
where, $\mathcal{R}(\varsigma)$ introduces a rotation matrix, and $\iota \in [0, \infty]$. The measurement $\mathcal{N}_b(\varepsilon)$ performed on Bob's mode $\delta b$ affects the probability distribution, defining a Gaussian measurement on Alice's mode $\delta a$. Interestingly, although Bob's Gaussian measurement yields an outcome $\varepsilon$ where $\mathcal{X}^{N_a,N_b}_{ab}=\mathcal{X}^{N_a}_a+\mathcal{C}^{N_b}$, it has no effect on the uncertainty in Alice's mode $\delta a$. $\mathcal{X}^{N_a}_a$ is given by Eq. \eqref{eq36}. The Shannon entropy of the relevant probability distribution, $H(\text{Pr}(\varepsilon,\eta))$, is comparable to the entropy of the Gaussian distribution $H(\mathcal{X}^{N_a,N_b}_{ab})$, which can be used to quantify the work extracted by Alice (optical mode). This is worded as \citep{Brunelli17}
\begin{equation}
\mathcal{W}^{(\upsilon,\iota)}(\theta,\varphi)=\frac{k_BT}{2}\ln\left(\frac{\det \mathcal{X}^{N_{b}}_{a}}{\det \mathcal{X}^{N_{a},N_{b}}_{ab}}\right).
\end{equation}
The extractable work corresponding to both homodyne ($\upsilon=0$) and heterodyne ($\upsilon=1$) measurements is given by \citep{33}
\begin{equation}
\mathcal{W}^{(0,0)}(\theta,\varphi)=\frac{k_BT}{2}\ln\left(\frac{4xy}{4xy-2z^2[1+2\cos(2\theta+2\varphi)]}\right),\quad \quad \mathcal{W}^{(1,1)}(\theta,\varphi)=k_BT\ln\left(\frac{(1+2x)(1+2y)}{1+2y+x(2+4y)-4z^2}\right).
\end{equation}

\section{Discussion and results} \label{sys1}
\subsection{First optomechanical system}

\begin{figure}[!h]
\centering
\begin{tabular}{ll}
\includegraphics[scale=1]{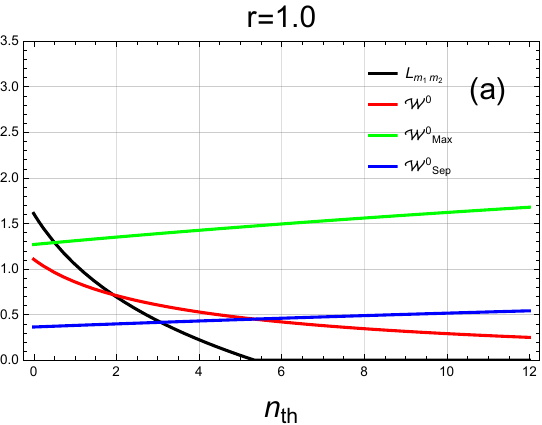} 
\includegraphics[scale=1]{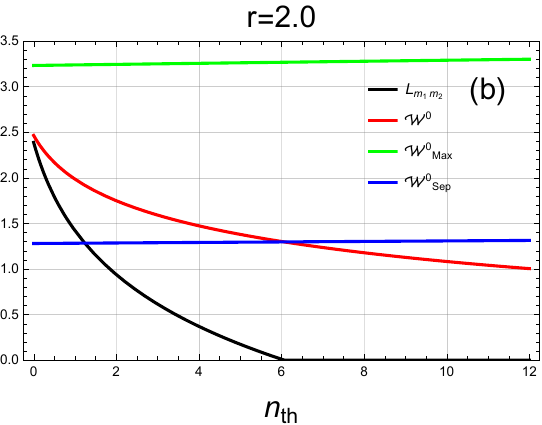} \\
\includegraphics[scale=1]{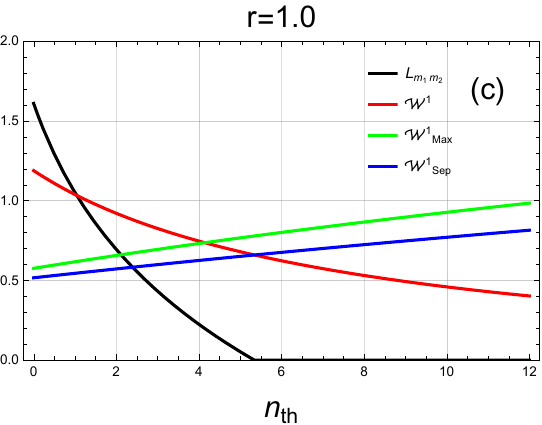} 
\includegraphics[scale=1]{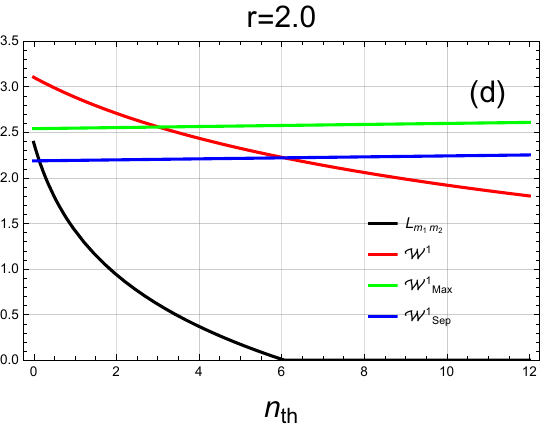}
\end{tabular}
\caption{Plot of the Logarithmic negativity $L_{m_1m_2}$ for mirror-mirror modes, the extractable work $\mathcal{W}^{(\upsilon)}$, the maximum work $\mathcal{W}_{\rm{Max}}^{(\upsilon)}$, and the separate work $\mathcal{W}_{\rm{Sep}}^{(\upsilon)}$ as functions of the thermal bath photon number $n_{th}$. These are shown for different values of the squeezing parameter $r$, with an optomechanical cooperativity of $C=34$ and $\frac{\Gamma}{\kappa}=0.05$. Panels (a) and (b) correspond to homodyne detection ($\upsilon=0$), while (c) and (d) correspond to heterodyne detection ($\upsilon=1$).}
\label{bath}
\end{figure}
Figure \ref{bath} shows the variation of mirror-mirror mode entanglement ($L_{m_1m_2}$), extractable work ($\mathcal{W}^{\upsilon}$), maximum work ($\mathcal{W}^{\upsilon}_{\rm{Max}}$), and separate work ($\mathcal{W}^{\upsilon}_{\rm{Sep}}$) as functions of the thermal bath photon number ($n_{th}$). These results are presented for various squeezing parameter values ($r$) in both homodyne ($\upsilon=0$) [panels a,b] and heterodyne ($\upsilon=1$) [panels c,d] measurements, with an optomechanical cooperativity of $C=34$ and $\frac{\Gamma}{\kappa}=0.05$. The figure clearly shows that both entanglement and extractable work rapidly diminish as $n_{th}$ (temperature $T$) increases. Notably, entanglement vanishes for a broad range of temperatures (or photon numbers $n_{th}$) \cite{28}. For a given $r$, the entanglement in the optical mode increases with increasing $r$. While an increase in temperature also contributes to the decrease of extractable work ($\mathcal{W}^{\upsilon}$), it remains non-zero across the entire range of $n_{th}$. Conversely, as $n_{th}$ increases, the mirror-mirror mode becomes separated ($\mathcal{W}^{\upsilon} < \mathcal{W}^{\upsilon}_{\rm{Sep}}$). Regarding the maximum work ($\mathcal{W}^{\upsilon}_{\rm{Max}}$), it increases with increasing temperature ($T$) for small squeezing parameter values ($r$). However, for larger $r$ values (specifically when $r>1$), the maximum work remains constant for all $T$ (or $n_{th}$), as depicted in Figure \ref{bath} (c) and (d). Furthermore, $\mathcal{W}^{\upsilon}_{\rm{Max}}$ generally increases with increasing $r$. For low photon numbers $n_{th}$, we observe that $\mathcal{W}^{1} > \mathcal{W}^{1}_{\rm{Sep}}$. However, beyond this regime, the separate work increases, and the extractable work $\mathcal{W}^{1}$ decreases. Finally, at $r=2$ and low $n_{th}$, we note $\mathcal{W}^{1} > \mathcal{W}^{1}_{\rm{Max}}$, with an overall increase in $\mathcal{W}^{\upsilon}$ as $r$ increases.
\begin{figure}[!h]
\centering
\begin{tabular}{ll}
\includegraphics[scale=1]{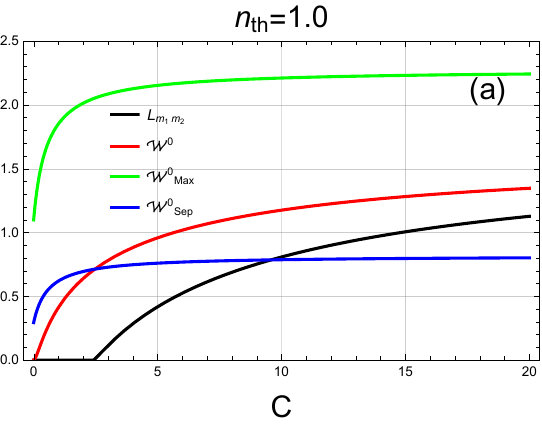} 
\includegraphics[scale=1]{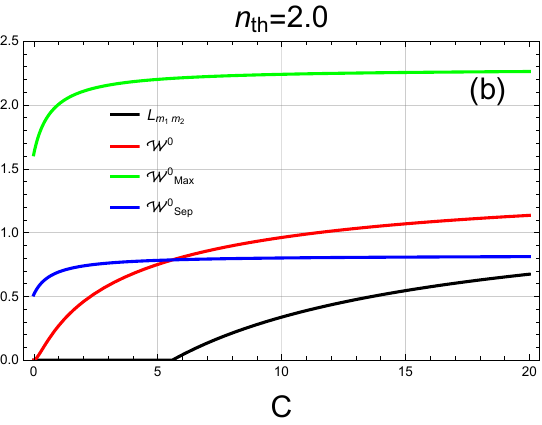} \\
\includegraphics[scale=1]{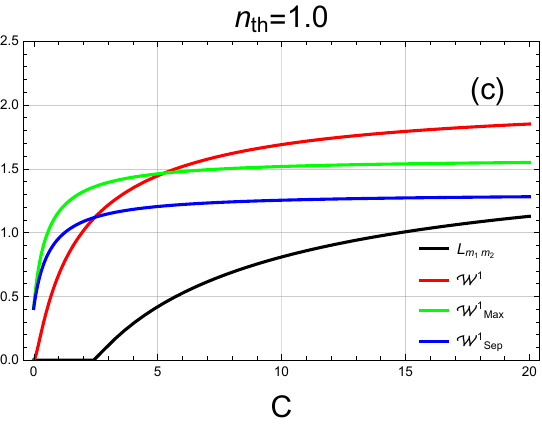} 
\includegraphics[scale=1]{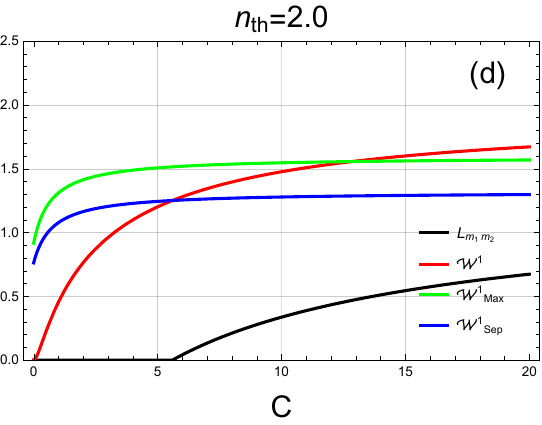} 
\end{tabular}
\caption{Logarithmic negativity ($L_{m_1m_2}$) between the two mirrors, extractable work ($\mathcal{W}^{\upsilon}$), maximum work ($\mathcal{W}_{\rm{Max}}^{\upsilon}$), and separate work ($\mathcal{W}_{\rm{Sep}}^{\upsilon}$) plotted as functions of cooperativity ($C$). These results are shown for thermal bath photon numbers $n_{th}=1.0$ and $n_{th}=2.0$, with a squeezing parameter $r=1.5$ and $\frac{\Gamma}{\kappa}=0.05$. Panels (a) and (b) correspond to homodyne detection ($\upsilon=0$), while (c) and (d) correspond to heterodyne detection ($\upsilon=1$).}
\label{Coop}
\end{figure}

We plot in Fig. \ref{Coop}, the logarithmic negativity ($L_{m_1m_2}$) and the three types of extractable work ($\mathcal{W}^{\upsilon}$, maximum extractable work $\mathcal{W}_{\rm{Max}}^{\upsilon}$, and separate work $\mathcal{W}_{\rm{Sep}}^{\upsilon}$) between two mirror modes, against the cooperativity ($C$). Results are shown for different values of the thermal bath photon number ($n_{th}$), or temperature ($T$), under homodyne detection ($\upsilon=0$) [panels a,b] and heterodyne detection ($\upsilon=1$) [panels c,d]. Remarkably, the condition of $C \neq 0$ is necessary to create entanglement between the two mirror modes. When the cooperativity exceeds a critical value, the entanglement ($L_{m_1m_2}$) increases. This critical value is dependent on the thermal bath photon number ($n_{th}$). Regarding the extractable work ($\mathcal{W}^{\upsilon}$), it increases for small cooperativity values but degrades as the thermal bath photon number ($n_{th}$) increases. The maximum work ($\mathcal{W}_{\rm{Max}}^0$) increases for small cooperativity values, and then remains constant as $C$ further increases. We also note that $\mathcal{W}^{1}>\mathcal{W}^{1}_{\rm{Max}}$ for large values of cooperativity. As the mean phonon number increases with augmentation of cooperativity, the mirror-mirror mode separates, meaning $\mathcal{W}_{\rm{Sep}}^{\upsilon}>L_{m_1m_2}$, as shown in Figure \ref{Coop}(b). The augmentation of the thermal bath phonon number also contributes to an increase in the separate work ($\mathcal{W}^{1}_{\rm{Sep}}$) and a decrease in the extractable work.
\begin{figure}[!h]
\centering
\begin{tabular}{ll}
\includegraphics[scale=1]{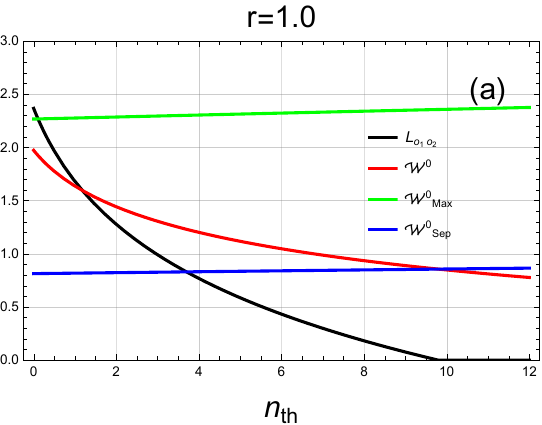} & 
\includegraphics[scale=1]{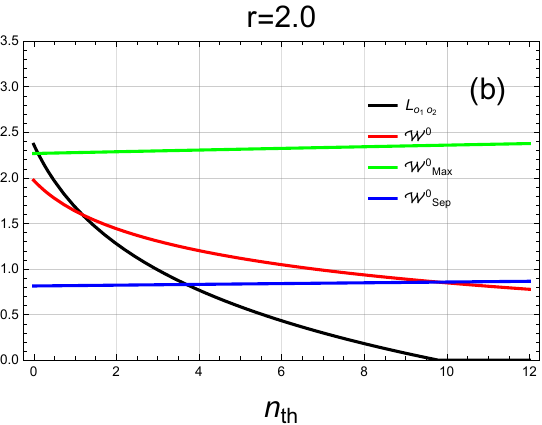}\\
\includegraphics[scale=1]{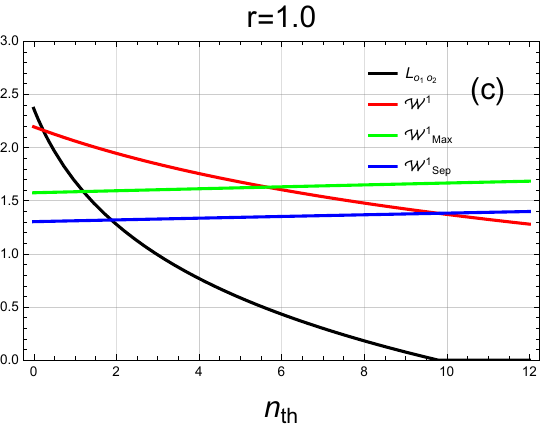} &
\includegraphics[scale=1]{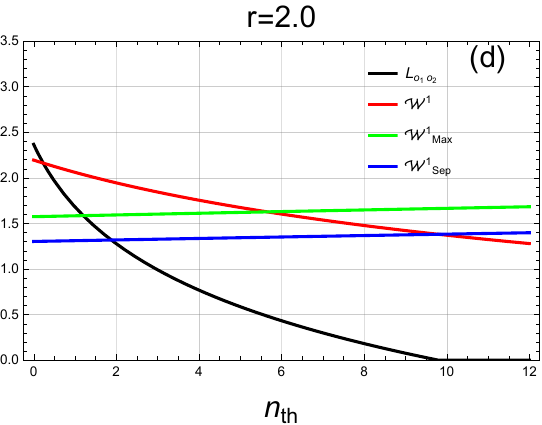}  
\end{tabular}
\caption{Plot of logarithmic negativity ($L_{o_1o_2}$) for optic-optic modes, extractable work ($\mathcal{W}^{\upsilon}$), maximum work ($\mathcal{W}_{\rm{Max}}^{\upsilon}$), and separate work ($\mathcal{W}_{\rm{Sep}}^{\upsilon}$) versus the thermal bath photon number ($n_{th}$). The data is presented for various squeezing parameter values ($r$), with $C=34$ and $\frac{\Gamma}{\kappa}=0.05$. Homodyne detection ($\upsilon=0$) is shown in panels (a) and (b), and heterodyne detection ($\upsilon=1$) in panels (c) and (d).}
\label{optic-optic}
\end{figure}

We explore in Fig.\ref{optic-optic}, the logarithmic negativity ($L_{o_1o_2}$) between two optic modes, as well as the extractable work ($\mathcal{W}^{\upsilon}$), maximum work ($\mathcal{W}^{\upsilon}_{\rm{Max}}$), and separate work ($\mathcal{W}^{\upsilon}_{\rm{Sep}}$) as functions of the thermal bath photon number ($n_{th}$) for various values of the squeezing parameter ($r$), under homodyne detection ($\upsilon=0$) [panels a,b] and heterodyne detection ($\upsilon=1$) [panels c,d]. The optomechanical cooperativity is set to $C=34$ and $\frac{\Gamma}{\kappa}=0.05$. This figure demonstrates that entanglement ($L_{o_1o_2}$) degrades with increasing $n_{th}$ and vanishes beyond a critical value, which is dependent on the squeezing parameter $r$. Notably, entanglement increases with increasing $r$. Conversely, the extractable work ($\mathcal{W}^{\upsilon}$) decreases as $n_{th}$ increases. For small values of $n_{th}$ and $r$, we observe $L_{o_1o_2}>\mathcal{W}^{\upsilon}$. As the squeezing parameter increases, the system quickly becomes separated ($\mathcal{W}^{\upsilon}_{\rm{Sep}}>L_{o_1o_2}$) for small $n_{th}$, and this separation remains constant for the entire range of thermal photon numbers. Regarding the maximum work ($\mathcal{W}^{\upsilon}_{\rm{Max}}$), we observe that it is consistently higher than the other works. Interestingly, the three extractable works remain non-zero even when the optic-optic mode is not entangled.

\begin{figure}[!h]
\centering
\begin{tabular}{ll}
\includegraphics[scale=1]{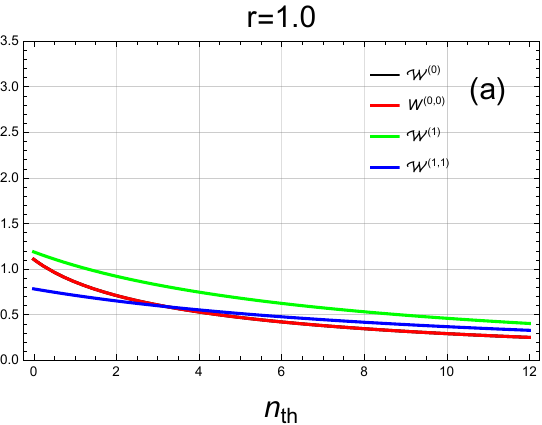} & 
\includegraphics[scale=1]{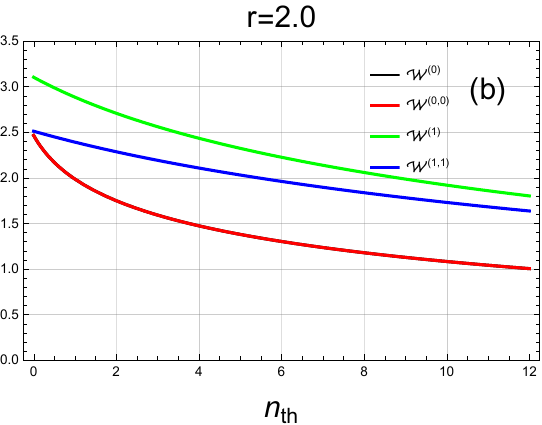}
\end{tabular}
\caption{Plot of the extractable work $\mathcal{W}^{(0)}$, work from two homodyne measurements $\mathcal{W}^{(0,0)}$, and work from two heterodyne measurements $\mathcal{W}^{(1,1)}$ between mirror-mirror modes, as a function of the thermal bath phonon number ($n_{th}$), for squeezing parameters (a) $r=1.0$ and (b) $r=2.0$, alongside results for single homodyne ($\mathcal{W}^{(0)}$) and heterodyne ($\mathcal{W}^{(1)}$) measurements.}
\label{figr7}
\end{figure}

Figure~\ref{figr7} compares the extracted work $\mathcal{W}^{(0,0)}$ and $\mathcal{W}^{(1,1)}$ for single homodyne ($\mathcal{W}^{(0)}$) and heterodyne ($\mathcal{W}^{(1)}$) measurements between two mirror modes as a function of thermal bath phonon number $n_{th}$, with squeezing parameters (a) $r=1.0$ and (b) $r=2.0$. We observe that: (i) the extracted work decreases with increasing $\mathsf{n}_{th}$, (ii) the work hierarchy reverses at large $n_{th}$ ($\mathcal{W}^{(1,1)} > \mathcal{W}^{(0,0)}$) compared to small $n_{th}$ ($\mathcal{W}^{(1,1)} < \mathcal{W}^{(0,0)}$), demonstrating temperature-dependent work reordering, and (iii) both $\mathcal{W}^{(0,0)}$ and $\mathcal{W}^{(1,1)}$ increase with the squeezing parameter $r$. We notice also that $\mathcal{W}^{(1)}>\mathcal{W}^{(1,1)}$ for whole values of the both parameters $n_{th}$ and $r$, and we observe that $\mathcal{W}^{(0,0)}=\mathcal{W}^{(0)}$ for whole values of the thermal bath photons numbers $n_{th}$ and squeeze parameter $r$.\\

\begin{figure}[!h]
\centering
\begin{tabular}{ll}
\includegraphics[scale=1]{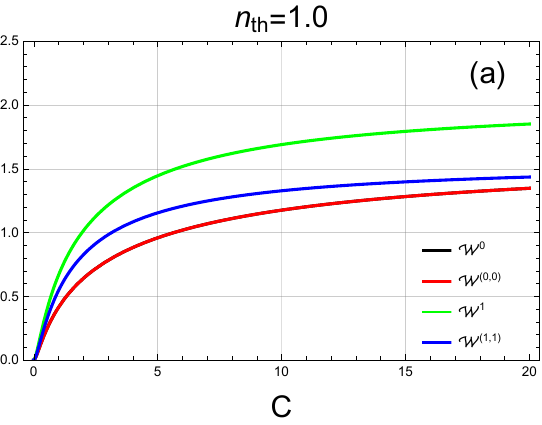} & 
\includegraphics[scale=1]{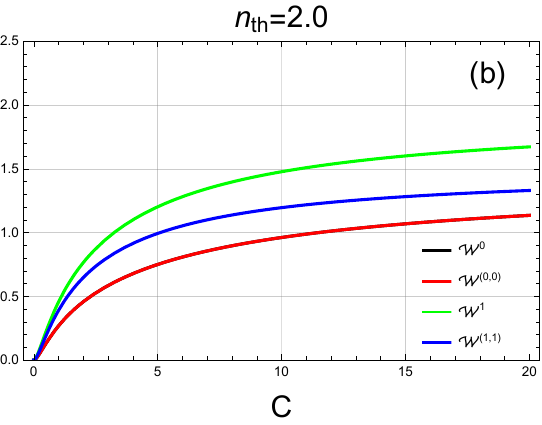}
\end{tabular}
\caption{Evolution of extracted work $\mathcal{W}^{\upsilon}$ and joint measurement works $\mathcal{W}^{(0,0)}$ and $\mathcal{W}^{(1,1)}$ for the mirror-mirror mode versus cooperativity $C$, with thermal bath phonon numbers (a) $n_{th}=1.0$ and (b) $n_{th}=2.0$. Results are shown for both single homodyne ($\mathcal{W}^{(0)}$) and heterodyne ($\mathcal{W}^{(1)}$) measurement protocols.}
\label{figr8}
\end{figure}

Figure~\ref{figr8}(a) shows the increase of extracted work $\mathcal{W}^{\upsilon}$ for both homodyne ($\upsilon=0$) and heterodyne ($\upsilon=1$) measurements, along with joint measurements $\mathcal{W}^{(0,0)}$ and $\mathcal{W}^{(1,1)}$ between two mirror modes, as cooperativity $C$ increases, where all work quantities vanish at $C=0$ and $\mathcal{W}^{(1)}$ exceeds $\mathcal{W}^{(1,1)}$ at large $C$, while Fig.~\ref{figr8}(b) demonstrates their decrease with thermal photon number $n_{th}$, consistent with Fig.~\ref{figr7}.

\begin{figure}[!h]
\centering
\begin{tabular}{ll}
\includegraphics[scale=1]{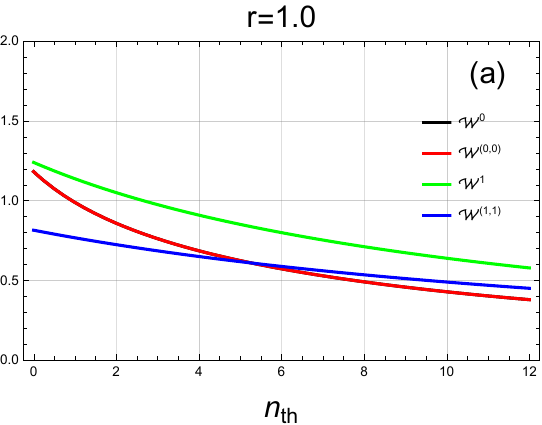} & 
\includegraphics[scale=1]{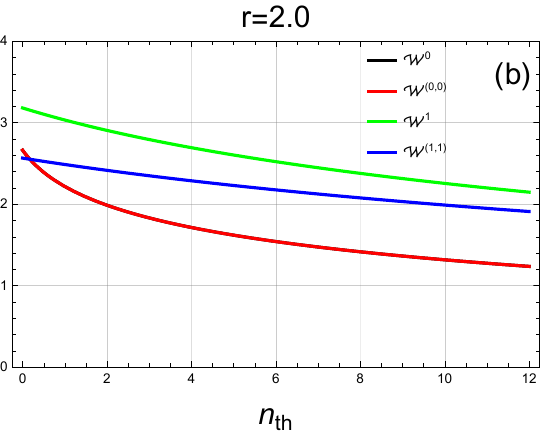}
\end{tabular}
\caption{Plot of the extract work $\mathcal{W}^{\upsilon}$ and both measurement $\mathcal{W}^{(0,0)}$ and $\mathcal{W}^{(1,1)}$ between the two optical modes versus the thermal bath phonons numbers $n_{th}$ for different values of the squeeze parameter (a) $r=1.0$ and (b) $r=2.0$, for both single homodyne measurement $\mathcal{W}^{(0)}$ and heterodyne measurement $\mathcal{W}^{(1)}$.}
\label{figr9}
\end{figure}

We compare the variation of the extractable work from both homodyne measurements ($\mathcal{W}^{(0,0)}$) and heterodyne measurements ($\mathcal{W}^{(1,1)}$) between two optic modes in Fig.\ref{figr9}. These are presented alongside the extractable work from a single homodyne measurement ($\mathcal{W}^{(0)}$) and a single heterodyne measurement ($\mathcal{W}^{(1)}$), all plotted as functions of the thermal bath phonon number ($n_{th}$). Results are shown for squeezing parameters (a) $r=1.0$ and (b) $r=2.0$. We observe that the extractable work decreases with increasing $n_{th}$. For large values of $n_{th}$, we find $\mathcal{W}^{(1,1)}>\mathcal{W}^{(0,0)}$, whereas for small values, $\mathcal{W}^{(1,1)}<\mathcal{W}^{(0,0)}$. This demonstrates the significant effect of temperature ($n_{th}$) on altering the relative order of these works. Furthermore, both $\mathcal{W}^{(0,0)}$ and $\mathcal{W}^{(1,1)}$ increase with increasing $r$.

\subsection{Second optomechanical system}

\label{sys2}
\begin{figure}[!h]
\centering
\begin{tabular}{ll}
\includegraphics[scale=1]{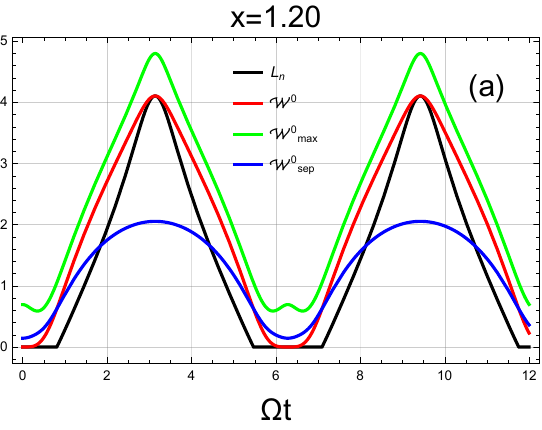}&
\includegraphics[scale=1]{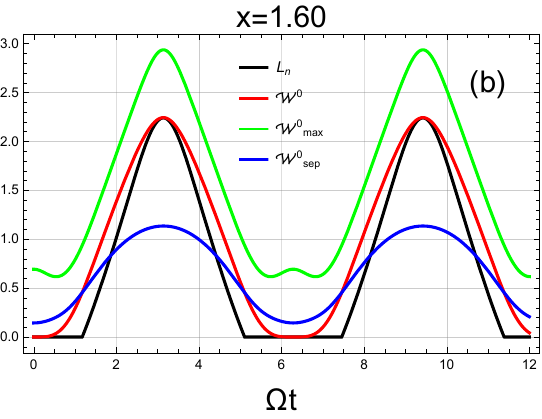}\\
\includegraphics[scale=1]{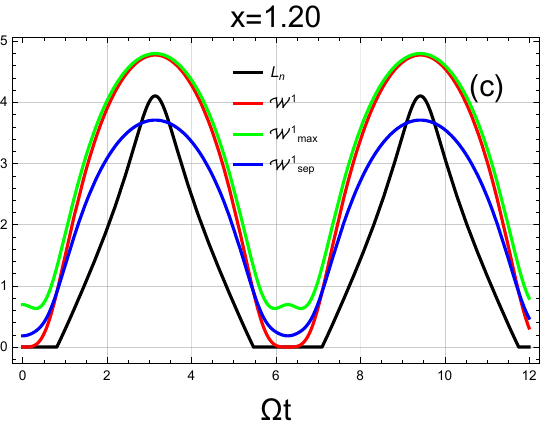}&
\includegraphics[scale=1]{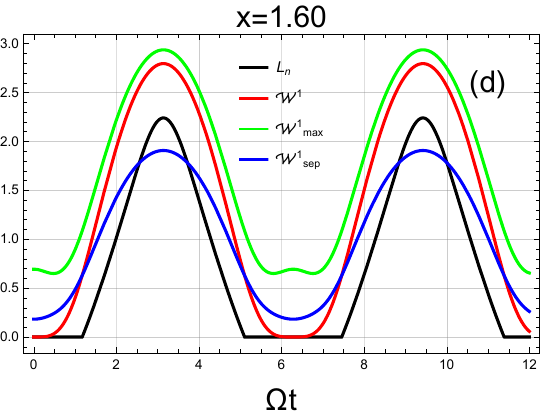}
\end{tabular}
\caption{Time evolution ($\Omega t$) of the logarithmic negativity $L_n$ optic-optic, extract work $\mathcal{W}^{\upsilon}$, maximum work $\mathcal{W}^{\upsilon}_{\rm{Max}}$ and separate work $\mathcal{W}^{\upsilon}_{\rm{Sep}}$ for various value of $x$, [a,b] $\upsilon=0$ (homodyne)[c,d], $\upsilon=1$ (heterodyne).}
\label{mo2}
\end{figure}

We present in Fig \ref{mo2}, the dynamical evolution of the logarithmic negativity ($L_n$) and extractable works ($\mathcal{W}^{\upsilon}$, maximum $\mathcal{W}^{\upsilon}_{\rm{Max}}$, and separate $\mathcal{W}^{\upsilon}_{\rm{Sep}}$) for the optic1–optic2 mode, for different values of $x$, under homodyne measurement $(\upsilon=0)$ and heterodyne measurement $(\upsilon=1)$. We remark that both the entanglement and the works vary periodically with time ($\Omega t$). Additionally, the peaks in Figure \ref{mo2} degrade more rapidly as the value of $x$ increases. When entanglement vanishes, we note that $\mathcal{W}^{\upsilon}_{\rm{Sep}}>\mathcal{W}^{\upsilon}$. This means that extracted work can be considered as a witness of entanglement. Furthermore, $\mathcal{W}^{\upsilon}_{\rm{Max}}>\mathcal{W}^{\upsilon}$ across the entire range of $\Omega t$. We also find that the separate work $\mathcal{W}^{1}_{\rm{Sep}}>\mathcal{W}^{0}_{\rm{Sep}}$ for the same value of $x$, and similarly for the extractable work $\mathcal{W}^{\upsilon}$. We notice that the time at which entanglement and extractable work $\mathcal{W}^{\upsilon}$ vanish increases with increasing $x$. This implies that the entanglement between optic1 and optic2 is significantly influenced by the value of $x$.

 \begin{figure}[!h]
\centering
\begin{tabular}{ll}
\includegraphics[scale=1]{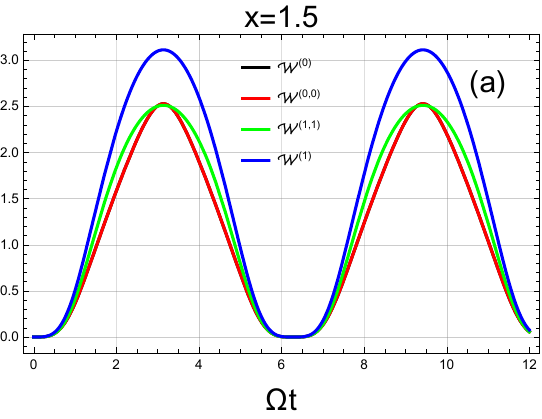} & 
\includegraphics[scale=1]{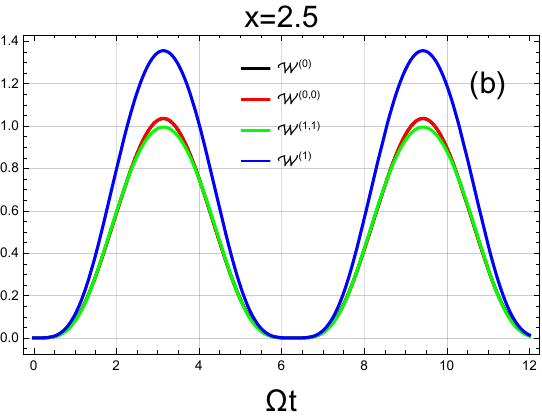}
\end{tabular}
\caption{Plot of the extractable work ($\mathcal{W}^{\upsilon}$), work from two homodyne measurements ($\mathcal{W}^{(0,0)}$), and work from two heterodyne measurements ($\mathcal{W}^{(1,1)}$) versus time ($\Omega t$). These results are shown for different values of the coupling ratio: (a) $x=1.5$ and (b) $x=2.5$. The figure also includes results for single homodyne measurement ($\mathcal{W}^{(0)}$) and single heterodyne measurement ($\mathcal{W}^{(1)}$).}
\label{figr11}
\end{figure}

In Figure \ref{figr11}, we show the dynamical evolution of the extractable work ($\mathcal{W}^{\upsilon}$), work from two homodyne measurements ($\mathcal{W}^{(0,0)}$), and work from two heterodyne measurements ($\mathcal{W}^{(1,1)}$) as functions of time ($\Omega t$). Results are presented for two values of the ratio: (a) $x=1.5$ and (b) $x=2.5$. We also include data for single homodyne ($\upsilon=0$) and heterodyne ($\upsilon=1$) measurements. We note the periodic temporal evolution of all four works. At their peaks, we find $\mathcal{W}^{(0,0)}>\mathcal{W}^{(1,1)}$, while $\mathcal{W}^{1}$ remains larger than the other works across the entire range of $\Omega t$. When the value of the ratio $x$ increases, all four works decrease. This indicates that an increasing $x$ leads to a reduction in the work obtained from both measurement types.

\section{Conclusion} \label{conc}

In summary, we have characterized quantum entanglement in relation with quantum thermodynamics in stationary and dynamical states of optomechanical systems. Our first system consists of two Fabry-P\'erot cavities coupled to a common two-mode squeezed light from the output of the parametric down conversion. In the second system we have examined the quantum correlations in the optomechanical system with a single vibrating mirror and a strong quasi-monochromatic laser incident on its surface. We have demonstrated that detecting of the entanglement between two modes is realized when $\mathcal{W}^{\upsilon}>\mathcal{W}^{\upsilon}_{\rm{Sep}}$ ($\upsilon=0,1$), separable and extracted works are always bounded by the maximum work in both measurement (homodyne and hetrodyne). Our findings indicate that all works are affected by various parameters such as the thermal bath photons number $n_{th}$, cooperativity $C$ and squeezed parameter $r$ for the first optomecanical system. We have shown also the variation of the both measurement various the same parameters.

\end{document}